\documentclass[amsmath,amssymb,aps,prb,reprint,floatfix,superscriptaddress]{revtex4-2}

\usepackage{booktabs}
\usepackage{enumitem}
\usepackage{graphicx}
\usepackage{glossaries}
\usepackage[version=4]{mhchem}
\usepackage[per-mode=symbol,separate-uncertainty=true]{siunitx}
\usepackage{hyperref}

\hypersetup{
    pdftitle={Competing Adsorption of H and CO on Pd-alloy Surfaces: Mechanistic insight into the mitigating effect of Cu on CO poisoning},
    pdfauthor={P. Ekborg-Tanner and P. Erhart},
    unicode,
    colorlinks=true,
    urlcolor=black,
    linkcolor=black,
    citecolor=black
}

\newacronym{ardr}{ARDR}{automatic relevance detection regression}
\newacronym{ce}{CE}{cluster expansion}
\newacronym{dft}{DFT}{density-functional theory}
\newacronym{fcc}{FCC}{face-centered cubic}
\newacronym{hcp}{HCP}{hexagonal close-packed}
\newacronym{lasso}{LASSO}{least absolute shrinkage and selection operator}
\newacronym{lspr}{LSPR}{localized surface plasmon resonance}
\newacronym{mae}{MAE}{mean absolute error}
\newacronym{mc}{MC}{Monte Carlo}
\newacronym{md}{MD}{molecular dynamics}
\newacronym{mlip}{MLIP}{machine-learned interatomic potential}
\newacronym{neb}{NEB}{nudged elastic band}
\newacronym{nep}{NEP}{neuroevolution potential}
\newacronym{p3fh}{P3FH}{pseudo 3-fold hollow}
\newacronym{rmse}{RMSE}{root mean square error}
\newacronym{sgc}{SGC}{semi-grand canonical}
\newacronym{spd}{SPD}{surface phase diagram}

\def\sectionautorefname{Sect.}
\def\figureautorefname{Fig.}

\def\equationautorefname{Eq.}

\newcommand{\Autoref}[1]{%
  \begingroup%
  \renewcommand\sectionautorefname{Section}%
  \renewcommand\subsectionautorefname{Section}%
  \renewcommand\figureautorefname{Figure}%
  \renewcommand\equationautorefname{Equation}%
  \autoref{#1}%
  \endgroup%
  }

\makeatletter
\let\oldtheequation\theequation
\renewcommand\tagform@[1]{\maketag@@@{\ignorespaces#1\unskip\@@italiccorr}}
\renewcommand\theequation{(\oldtheequation)}
\makeatother

\DeclareSIUnit\angstrom{\text{Å}}
\DeclareSIUnit\atom{\text{atom}}
\DeclareSIUnit\bar{\text{bar}}
\DeclareSIUnit\step{\text{step}}
\DeclareSIUnit\Matom{\text{metal atom}}

\newcolumntype{d}{D{.}{.}{-1}}

\begin{document}

\title{
    Competing adsorption of H and CO on Pd-alloy surfaces:\texorpdfstring{\\}{}
    Mechanistic insight into the mitigating effect of Cu on CO poisoning
}

\author{Pernilla Ekborg-Tanner}
\author{Paul Erhart}
\email{erhart@chalmers.se}
\affiliation{Department of Physics,
    Chalmers University of Technology,
    SE 412~96 Gothenburg, Sweden}

\begin{abstract}
Multi-component alloys offer broad tunability for addressing challenges in materials science, but their vast configurational space makes their surface chemistry highly sensitive to operating conditions, for example through adsorption and segregation. 
Here, we study Pd--Au--Cu alloy surfaces in \ce{H2} and \ce{CO} environments motivated by their use in H technologies, in particular plasmonic \ce{H2} sensing, where alloying can mitigate limitations intrinsic to Pd such as hysteresis and CO poisoning. 
Modeling multicomponent surfaces with multiple adsorbate species under realistic conditions is challenging. 
To this end, we establish an accurate and efficient framework that combines \glsentrylongpl{mlip} trained on \glsentrylong{dft} data to generate training data for \glsentrylongpl{ce} with effectively no limitations on training set size.

By constructing continuous surface phase diagrams for H--CO coadsorption we find that coadsorption under operating conditions is governed primarily by the H coverage during annealing. 
Au-rich surfaces, formed under H-poor conditions, suppress both CO and H adsorption, while H-rich conditions yield Pd-rich surfaces that maintain higher H coverages compared to Pd at relevant CO partial pressures, indicating improved CO poisoning resistance. 
This effect is insensitive to relative amounts of Au and Cu, despite experimental evidence of the mitigating effect of specifically Cu on CO poisoning. 
Kinetic barriers for dilute alloy surfaces indicate that absorption pathways near Au are highly unfavorable, while Cu leave the energetics unchanged compared to pure Pd. 
This finding suggests that Cu in the surface region provides viable pathways to shuttle H into the material when Pd-dominated paths are blocked by CO.
\end{abstract}

\maketitle

\section{Introduction}

Multi-component materials underpin applications ranging from chemical sensing and catalysis to energy conversion and structural engineering.
Under realistic operating conditions, such as finite temperatures and reactive gas environments, the surface properties of these materials are often changed due to adsorption and segregation \cite{LovOpa08, EkbErh21}. 
Such processes critically influence materials performance by altering kinetic, thermodynamic, electronic, optical, and mechanical properties.
Understanding and controlling these phenomena is, therefore, essential for the design of functional multi-component materials.
Yet, current modeling approaches still struggle to capture the immense combinatorial and structural complexity of these systems, particularly when adsorbates, surfaces, and thermodynamic effects all come into play.

Pd-based nanoalloys have received considerable attention, especially in the context of optical hydrogen sensing, which exploits changes in the position and shape of the \gls{lspr} of Pd nanoparticles upon hydrogen sorption \cite{LanLarKas10}.
While the basic working principle is well established, further improvements are needed to make this technology robust under realistic conditions.
Pure Pd sensors suffer from hysteresis between hydrogen absorption and desorption, as well as carbon monoxide (CO) poisoning caused by the preferential adsorption of CO over H \cite{WelSilGar15}.
These effects directly compromise the reliability and stability of \gls{lspr}-based hydrogen sensors.
Alloying is an effective mitigation strategy, as the addition of approximately \qty{25}{\percent} Au suppresses hysteresis, while introducing only \qty{5}{\percent} Cu has been found to reduce CO poisoning \cite{WadNugLid15, DarNugKad19, DarKhaTom21}.
While the effect of Au on hysteresis is well understood \cite{RahLofFra21}, the mechanism behind the mitigating effect of Cu is not yet understood, especially since Cu rarely resides directly at the surface \cite{EkbErh21}. 

Introducing multiple alloying elements and adsorbate species, however, substantially enlarges the design space and creates additional complexity due to the nontrivial coupling between adsorption and segregation.
This system thus provides a technologically relevant and scientifically rich case that demands advancing models of multi-component materials under reactive conditions.

To comprehensively model such a system, a framework is required that (i) captures subtle energy differences between competing atomic arrangements in the bulk and at surfaces, (ii) treats the interactions among adsorbates and between adsorbates and surfaces with sufficient accuracy, and (iii) enables efficient sampling across a range of thermodynamic boundary conditions, including temperature and gas pressure.
Here, we address these challenges by combining \glspl{mlip} trained on \gls{dft} calculations with \glspl{ce} and \gls{mc} sampling and apply this approach to study the competing adsorption between H and CO on PdAuCu alloy surfaces. 

We find that the surface coverage of H and CO depends on the chemical configuration of the surface region, which is determined by the preparation conditions.
In addition, we find that alloying increases the H-to-CO ratio compared to pure Pd, regardless of the exact alloy composition.  
The adsorption thermodynamics alone does not, however, explain the experimental observation that Cu is necessary to mitigate CO poisoning. 
To gain further insight into the CO poisoning mechanism, we therefore study the kinetics of H sorption into the bulk by analyzing the kinetic barriers associated with H absorption of dilute PdAu and PdCu alloys. 
We find that paths close to Au are highly unfavorable, while Cu displays similar energetics as pure Pd. 
This leads us to propose that Cu mitigates CO poisoning by providing viable pathways in situations where the most favorable paths are blocked by CO.

\section{Methodology}
\label{sect:approach}

Electronic-structure calculations, most commonly based on \gls{dft}, remain the common standard for accuracy when modeling extended systems.
However, their high computational cost renders them impractical for systems with multiple alloying components, low symmetries, and finite-temperature effects such as mixing, ordering, and segregation.

\Glspl{ce} provide an efficient and well-established framework for modeling configurational complexity in alloys \cite{SanDucGra84}.
They have been successfully applied to multi-component bulk alloys \cite{BeaWatVal24} as well as surfaces with adsorbates \cite{EkbErh21, MulSto06, MueCed09}.
In low-symmetry systems, such as surfaces, the number of interaction parameters grows rapidly, raising the demand for training data, which is still most commonly generated using \gls{dft} calculations.
While Bayesian approaches help manage this growth by incorporating physical insight via priors \cite{MueCed09}, the combinatorial explosion that occurs when increasing the number of components remains challenging \cite{BeaWatVal24, ZhaLiuBi20, MulNat25}.
When multiple surface orientations are relevant, it is furthermore desirable for the different surface-specific \glspl{ce} to yield consistent behavior in the bulk limit.
Without such consistency, predictions of surface segregation across facets may become unreliable, which poses the additional challenge of reconciling models for different surface orientations.

In the last decade, \glspl{mlip} have emerged as powerful tools that bridge the accuracy of electronic-structure methods with the efficiency of empirical potentials.
They open the possibility of serving as intermediaries for constructing \glspl{ce}, offering accuracy at scale \cite{ShoHolDas24}.
While \glspl{mlip} have already been deployed in hybrid \gls{md}-\gls{mc} simulations to model ordering and mixing in bulk alloys \cite{SadErhStu12, SonZhaLiu24}, these approaches are not well suited for sampling adsorbate--gas equilibria.
For this purpose, lattice-based \gls{ce}-\gls{mc} simulations remain the most effective strategy.

To tackle the above challenges, we adopt the following approach.
To cut the cost of \gls{dft} calculations, we build \glspl{mlip} based on both the \gls{nep} and the MACE architecture.
Specifically, we use \gls{nep} models, which are up to three orders of magnitude faster than MACE (\autoref{sfig:timing-comparison}), to build a comprehensive set of reference data via active learning, and MACE models, which yield higher accuracy, to generate reference data for training \gls{ce} models.
The \glspl{ce} are constructed for \{111\}, \{110\}, and \{100\} surfaces that are coupled to the same underlying bulk model through constraints, and include Au, Pd, and Cu on the metal lattice as well as H and CO on the adsorbate lattices.
Training data for these models are generated from structural relaxation using the MACE model, such that relaxation effects are implicitly included, in an active learning loop.
In taking this approach, we exploit that different types of predictions require different levels of accuracy.

\begin{figure}
\centering
\includegraphics[]{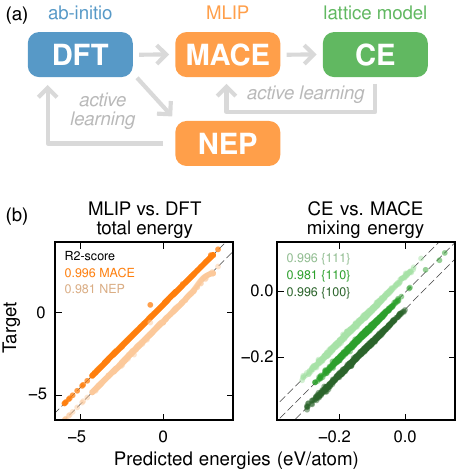}
\caption{
\textbf{Model development.}
    (a) Schematic illustration of the computational framework.
    (b) Model errors for the \glspl{mlip} and the \gls{ce}. 
    Note that the data has been systematically shifted along the y-axis to allow for clearer visualization of the different models. 
}
\label{fig:mlip-training}
\end{figure}

\begin{figure}[t]
\centering
\includegraphics[scale=0.92]{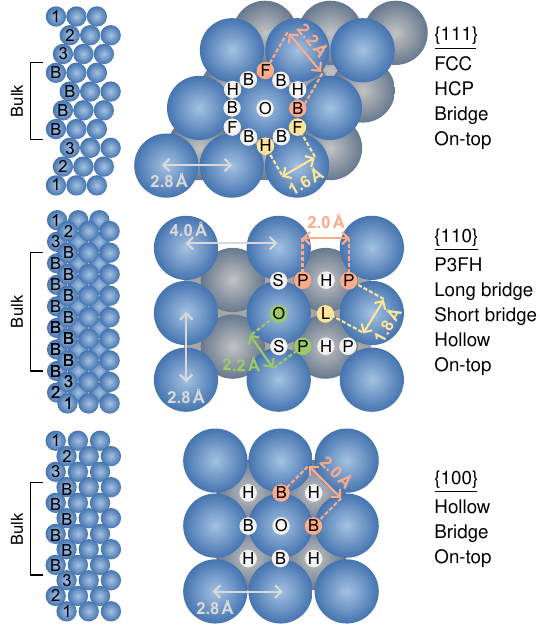}
\caption{
    \textbf{Schematic illustration of the surface slabs and their respective adsorption sites.}
    Side views of the surface slabs (to the left) with the surface layers and bulk region indicated. 
    Top views of the surfaces (to the right) with
    blue atoms for the top layer, gray atoms for the subsequent layers.
    The smaller atoms show the adsorption sites labeled by their first letter. 
    The arrows between sites indicate their three-dimensional assuming a \qty{4.0}{\angstrom} lattice parameter. 
}
\label{fig:ads-sites}
\end{figure}

\begin{figure*}
\centering
\includegraphics{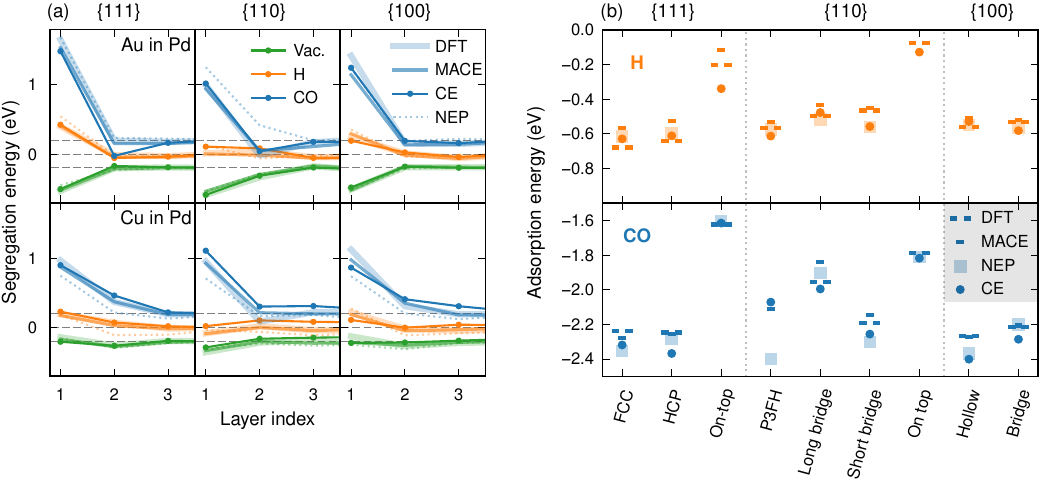}
\caption{
    \textbf{Model performance for surface properties.}
    (a) Segregation energies from \gls{dft}, \gls{nep}, MACE, and \gls{ce} for dilute Pd alloys in vacuum, \ce{H2} and CO (\qty{100}{\percent} coverage).
    Negative (positive) values indicate that the minority species prefers (avoids) residing in the surface region.
    Note that the curves are shifted along the y-axis for better visualization. 
    (b) Adsorption energies for H and CO on Pd (\qty{25}{\percent} coverage) for all models.
    Additional sites and adsorption energies on Cu and Au can be found in \autoref{sfig:adsorption-energies-all}.
}
\label{fig:model-performance}
\end{figure*}

\subsection{Machine-learned interatomic potentials}

We construct \glspl{mlip} using the fourth-generation \gls{nep} framework \cite{SonZhaLiu24, FanWanYin22}, the separable neuroevolution strategy \cite{WieSchGla14} and the iterative procedure described in Ref.~\citenum{FraWikErh23}, utilizing the \textsc{gpumd} \cite{FanWanYin22, XuBuPan25} and \textsc{calorine} \cite{LinRahFra24} packages, as detailed in \autoref{snote:nep}.
A combination of structure enumeration \cite{HarFor08, HarFor09}, \gls{neb} \cite{HenJon00, HenUbeJon00, LinKasPet19} pathways and \gls{md} trajectories are used to generate bulk and surface structures in an active training loop, using the \gls{nep} models and the \textsc{ase} \cite{LarMorBlo17}, \textsc{hiphive} \cite{EriFraErh19}, and \textsc{icet} packages \cite{AngMunRah19, EkbRosFra24}. 
The final set of reference structures comprises \num{18403} structures, corresponding to a total of \num{586006} atoms (see \autoref{stab:training-set-nep} for an overview).
Using the final set of reference structures, we construct a MACE model \cite{BatKovSim22} using the \textsc{mace-torch} package \cite{BatBatKov22} as described in \autoref{snote:mace}.

\subsection{Density functional theory calculations}

The \glspl{mlip} are fitted using forces, energies, and virials obtained from \gls{dft} calculations, employing the van-der-Waals density functional with the consistent exchange (vdW-DF-cx) method \cite{DioRydSch04, BerHyl14} as implemented in the Vienna ab-initio simulation package \cite{KreFur96a, KreFur96b} using projector-augmented wave setups \cite{Blo94, KreJou99} with a plane wave energy cutoff of \qty{520}{\electronvolt}.
The Brillouin zone is sampled with automatically generated $\mathbf{k}$-point grids with a maximum spacing of \qty{0.1}{\per\angstrom}.

\subsection{Cluster expansion construction}

\Glspl{ce} are constructed for \gls{fcc} surface-adsorbate slabs with \{111\}, \{110\} and \{100\} orientation using the \textsc{icet} package \cite{AngMunRah19} following the procedure outlined in Ref.~\citenum{EkbRosFra24}.
The \glspl{ce} predict the mixing energy based on training data obtained by atomic relaxation and energy calculation using the MACE model. 
The mixing energies are calculated by using the energy per atom of a pure metal slab for the metal species and the molecular ground state energy (divided by 2 for \ce{H2}) for the adsorbate species.
\Autoref{fig:ads-sites} shows all atomic sites of the slab and adsorbates. 
The number of layers varies between the slabs to achieve a similar slab thickness, with 10 layers for \{111\}, 18 for \{110\} and 12 for \{100\}.

The adsorbate sublattice represents an open system and can be occupied by H, CO, or vacancies.
All relevant adsorption sites are considered, which allows for configurations with coverages far beyond the relevant limit.
Since we do not take into account H sorption into the bulk, only surface coverages up to the limit where entering the subsurface becomes favorable are of interest.
This corresponds to a 1:1 ratio between adsorbate and surface metal atoms for \{111\} and \{100\} and 2:1 for \{110\} (\autoref{sfig:Pd-ads-abs}). 
The larger allowed coverage on \{110\} surfaces is a consequence of the large surface area of \{110\} where almost the entire second atomic layer is exposed (\autoref{fig:ads-sites}). 
For this reason, we define \qty{100}{\percent} surface coverage on \{110\} to mean a 2:1 ratio, as opposed to 1:1 for the other facets, which allows for a fairer comparison between the surface orientations.

Cutoffs are selected to include pairs and triplets up to distances of \qty{8.0}{\angstrom} and \qty{5.0}{\angstrom}, respectively, for a \qty{4.0}{\angstrom} lattice parameter.
The number of interaction parameters depends on the number of layers and adsorption sites and ranges from \num{3300} (\{100\}) to \num{9600} (\{110\}) for the systems considered here. 
In the context of \gls{ce} models, these are very large parameter spaces which would be difficult to fit properly using conventional approaches.
To overcome this challenge, we exploit pruning and merging of clusters.

Because the adsorbate lattice contains many close-lying, symmetrically inequivalent sites, their interactions will con\-tri\-bute to a very large number of inequivalent clusters. 
Many of these correspond to interaction lengths shorter than the minimal distance between adsorbates at the maximum coverage limit. 
For this reason, we introduce a lower cutoff radius of \qty{1.5}{\angstrom} for \{111\} and \{100\}, and \qty{1.2}{\angstrom} for \{110\} and prune all clusters below this limit.
We also make the pragmatic choice of pruning all adsorbate-adsorbate interactions longer than the \gls{fcc} nearest-neighbor distance (\qty{2.8}{\angstrom}) as well as all adsorbate-slab interactions extending below the third layer. 

Merging, in this context, refers to grouping together clusters that are not strictly symmetrically equivalent but can be assumed to be similar \cite{EkbErh21, EkbRosFra24}.
This includes clusters far from the surface region, which should be similar to the equivalent cluster of a bulk system. 
Here, we consider the three outermost atomic layers to be the surface region and all inner layers to be the bulk, and merge all clusters with the same radius that consist of bulk sites only, as described in Ref.~\citenum{EkbRosFra24}.
To ensure the \gls{ce} models for the different surface orientations yield a correct and consistent description of the bulk region, we first construct a bulk \gls{ce} for the AuCuPd system based on approximately 900 enumerated structures with up to 9 atoms in the unit cell and up to \qty{50}{\percent} Au and \qty{20}{\percent} Cu.
The resulting bulk interaction parameters are used in place of the merged bulk interactions for the surface slabs. 

Pruning, merging and fixing the bulk interactions cuts the number of orbits by half, resulting in \numrange{2000}{4500} parameters, which can be further reduced by regularization. 
Because these numbers are still too large to efficiently apply common advanced linear regression methods, a two step approach is adapted. 
First, an initial feature selection is performed using the \gls{lasso} \cite{Tib96} resulting in \numrange{1000}{1400} parameters. 
Second, \gls{ardr} \cite{Mac94} is used to fit the final model after tuning the hyperparameters to achieve the lowest number of features with a converged R2-score (\autoref{sfig:ce-learning-curves}).
We use \gls{lasso} and \gls{ardr} as implemented in \textsc{scikit-learn} \cite{PedVarGra11} via the \textsc{trainstation} package \cite{FraEriErh20}.

Atomic structures for the training set are generated and selected using an active learning approach and relaxed using the MACE model, keeping the in-plane lattice parameter fixed based on interpolation of lattice parameters from relaxed bulk structures. 
An initial dataset, consisting of \num{300} structures with approximately orthogonal cluster vectors (following the method outlined in Ref.~\citenum{EkbRosFra24}), is used to train an ensemble of \glspl{ce}.
A \gls{ce} corresponding to the mean of the ensemble is used to run \num{300} short annealing \gls{mc} simulations.
The ensemble of \glspl{ce} is used to calculate the uncertainty in energy across these trajectories, and the structure with the highest uncertainty for each trajectory is added to the training set. 
This process is repeated four times, at which point the \gls{rmse} is no longer improving (\autoref{sfig:ce-learning-curves}). 

For the final models, we also include the structures generated specifically for calculating the adsorption and segregation energies as well as similar sets of structures generated for adsorbate-free \glspl{ce} in parallel, resulting in about \num{2800} structures for each orientation.
The final models show excellent agreement with the MACE training data (\autoref{fig:mlip-training}b).
\sloppy For the \{111\} surface, the final model has \num{323} features (i.e., non-zero parameters) and achieves a validation \gls{rmse} of \qty{3.8}{\milli\electronvolt\per \Matom}, for the \{110\} surface the final model has \num{281} features and achieves a validation \gls{rmse} of \qty{6.4}{\milli\electronvolt\per\Matom}, and finally, for the \{100\} surface the final mode has \num{195} features and yields a validation \gls{rmse} of \qty{3.7}{\milli\electronvolt\per\Matom}.
The models are further validated by comparing the predicted segregation and adsorption energies with \gls{dft} and \gls{mlip} data (\autoref{fig:model-performance}), which demostrates excellent agreement. 
Here, the segregation energy is calculated for a single Au (Cu) atom in different layers of a $2\times2$ Pd slab, compared to the energy of placing the minority atom in the bulk.

\subsection{Monte Carlo simulations}\label{sect:mc}

To study the surface-adsorbate system under various conditions, \gls{mc} simulations are performed using the \textsc{mchammer} package \cite{AngMunRah19}.
The structures studied typically consist of $6\times6 \times n_\text{layers}$ metal atoms and up to 1 (for \{111\} and \{100\}) or 2 (for \{110\}) monolayers of H or CO and the simulation runs for 100 \gls{mc} cycles. 

\Gls{mc} simulations of an adsorbate lattice need to account for the fact that many unfavorable configurations will not be represented well by the \gls{ce}, since the adsorbates can move to more favorable configurations during relaxation. 
Especially when using the full adsorbate lattice (i.e., all adsorption sites), distances between neighboring sites are typically too short to allow for occupation of both sites in atomic relaxation as well as in reality. 
To account for this, we impose a minimal distance between occupied adsorbate sites (\qty{1.9}{\angstrom} for \{111\} and \{110\}, \qty{2.1}{\angstrom} for \{100\} \autoref{fig:ads-sites}), which still allows for reaching the maximum coverage before absorption to the subsurface becomes favorable (\autoref{sfig:Pd-ads-abs}).
In addition, adsorbate sites not well represented by the \glspl{ce} (bridge for \{111\}, hollow for \{110\} and on-top for \{100\}, see \autoref{sfig:adsorption-energies-all}) are excluded in the \gls{mc} simulations.

The \gls{mc} simulations are performed in multiple steps to mim\-ic the experimentally studied systems, which were annealed in a \ce{H2} environment (corresponding to \qty{40}{\milli\bar} H partial pressure) at \qty{773}{\kelvin} before the optical measurements were performed in ambient conditions and varying \ce{H2} and \ce{CO} partial pressures. 
First, the slab configuration is obtained from \gls{mc} simulations at \qty{773}{\kelvin} and a varying H coverage. 
Here, separate canonical ensembles are used for the bulk, surface and adsorbate subsystems.
The equilibrium composition of the surface region, in relation to the bulk and adsorbates, is found by interpolating results from preceding simulations where bulk and surface are treated as one subsystem. 
Then, the co-adsorption phenomena during the experimental measurements is simulated via \gls{mc} simulations at \qty{300}{\kelvin}.
Due to the different timescales of adsorption and chemical reordering of the slab configuration, were the latter is much slower, we keep the slab configuration fixed (from the previous simulation) while the adsorbate sublattices equilibrates within a \gls{sgc} ensemble controlled by the H and CO chemical potentials. 

To prevent the adsorbate configuration from freezing in, swaps in the \gls{sgc} ensemble are mixed with swaps in a canonical ensemble for the adsorbate sublattice, which enables swaps between two sites. 
This scheme enables moving an adsorbate from, e.g., a bridge site to a neighboring \gls{fcc} site without having to first switch the site to vacuum which rarely will be an accepted move.

\section{Results and discussion}

\subsection{Surface phase diagram construction}

\begin{figure}
\centering
\includegraphics[scale=0.97]{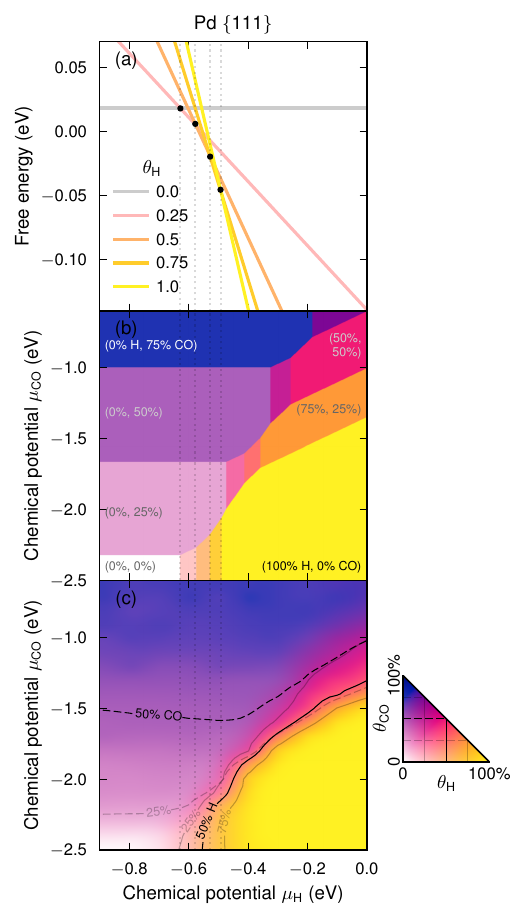}
\caption{
    \textbf{\gls{spd} construction for Pd \{111\}.}
    Traditionally, \glspl{spd} are constructed by calculating the free energy as a function of the chemical potential for specific coverages (a) and translating the convex hull to a discrete surface diagram (b). 
    \Gls{mc} sampling of \glspl{ce} allows for efficient sampling of surface coverage(s) as a function of chemical potential(s) which results in continuous surface diagrams (c). 
}
\label{fig:spd-construction}
\end{figure}

Traditionally, \glspl{spd} are constructed by calculating the free energy per primitive cell (or area) as a function of chemical potential at a few selected surface coverages and identifying the minimum energy phase.
In \autoref{fig:spd-construction}, we show this process for the Pd \{111\} surface.
In \autoref{fig:spd-construction}a, the free energy in the \qty{0}{\kelvin} limit,
\begin{equation}
    G(\theta_\text{H}, \theta_\text{CO} = 0) = E^\text{mix}(\theta_\text{H}, \theta_\text{CO}) - \theta_\text{H} \mu_\text{H},
\end{equation}
is calculated from the \gls{ce} mixing energy (from which the gas phase \ce{H} and CO energies in relation to their respective coverages $\theta$ have already been subtracted) and the chemical potential. 
The convex hull indicates the stable adsorbate coverage in a certain chemical potential interval. 
By repeating this for multiple combinations of H and CO coverages, a two-dimensional \gls{spd} can be constructed (\autoref{fig:spd-construction}b). 
While this example uses a \gls{ce}, it is straightforward to construct this kind of surface diagram using \gls{dft} calculations for pure metal surfaces and certain high-symmetry alloys (see \autoref{snote:SPD}, \autoref{sfig:SPD-DFT-CE}).
For more complex system, however, the number of configurations to consider quickly becomes unmanageable.
In addition, finite temperatures require treatment of the configurational entropy. 
Using a \gls{ce} in combination with \gls{mc} simulations allows for efficient thermodynamic sampling, including configurational entropy contributions, in the \gls{sgc} ensemble where the chemical potentials are controlled, which results in a continuous \gls{spd} (\autoref{fig:spd-construction}c).

\subsection{Pressure conversion}

\begin{figure*}[t]
\centering
\includegraphics[]{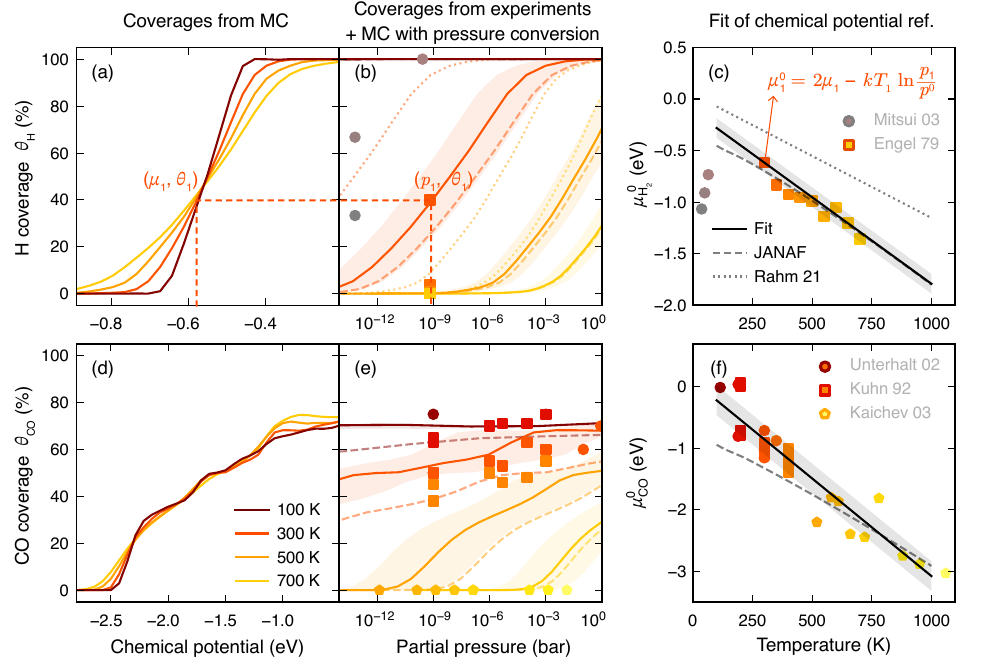} 
\caption{
    \textbf{Conversion between chemical potential and pressure for Pd.}
    (a) H coverage vs. chemical potential obtained from \gls{mc} simulations. 
    (b) Experimental records of H coverage vs. partial pressure from Mitsui 03 \cite{MitRosFom03}, Engel 79 \cite{EngKui79}, and additional sources (see \autoref{sfig:pressure-conv}) \cite{WilMatFuk01}.
    (c) Fit of the reference chemical potential based on the data in (a) and (b) as well as the calculated reference chemical potential based on tabulated thermodynamic data (JANAF \cite{nist-janaf}) or computational models (Rahm 21 \cite{RahLofFra21}).
    Once the reference chemical potential is known, the \gls{mc} data (a) can be converted from chemical potential to partial pressure, as shown in (b) for the different references. 
    (d--f) show the same process for CO with experimental data for Pd:CO from Unterhalt 02 \cite{UntRupFre02}, Kuhn 92 \cite{KuhSzaGoo92}, Kaichev 03 \cite{KaiProBuk03}, and additional sources (see \autoref{sfig:pressure-conv}) \cite{HeMemGri88, HeNor88, BehChrErt80}.
    Note that the legends are shared across each row. 
    }
\label{fig:pressure-conversion}
\end{figure*}

In most cases, the partial pressure $p$ is more relevant than the chemical potential $\mu$. 
If ideal gas behavior can be assumed, the two are connected via the relations
\begin{align}
    \mu_\text{H} (T, p) &= \frac{1}{2}\left(\mu^0_{\text{H}_2} (T) + k T \ln\frac{p_{\text{H}_2}}{p^0}\right) \\
    \mu_\text{CO} (T, p) &= \mu^0_\text{CO}(T) + k T \ln\frac{p_\text{CO}}{p^0}
    \label{eq:pressure-conversion}
\end{align}
for \ce{H} and \ce{CO}, respectively.
The reference chemical potential $\mu^0(T)$, defined at a reference partial pressure $p^0$, can, in principle, be obtained from thermodynamic tables \cite{nist-janaf}.
This approach, however, suffers from the fact that adsorption energies are generally associated with a significant error in \gls{dft} calculations, leading to an effective shift of the chemical potential. 
In the context of our modeling framework, this approach suffers from the fact that the temperature dependence of the adsorbed molecules is limited to configurational entropy, while the full temperature dependence is accounted for with regard to the gas phase.

In some cases, one can find an observable that can be measured experimentally and calculated in the modeling framework and use this as a basis for calculating the reference. 
In the present work, the adsorbate coverage at a given temperature can be used for this purpose by linking the coverage as a function of chemical potential from \gls{mc} simulations to experimental observations of coverage as a function of partial pressure.  
With this approach, the full temperature dependence is implicitly taken into account, but the availability of experimental data at relevant conditions as well as uncertainty in the measurements are potential issues. 

In \autoref{fig:pressure-conversion} we show our approach to pressure conversion. 
The H coverage on pure Pd is obtained as a function of the H chemical potential (\autoref{fig:pressure-conversion}a) from \gls{mc} simulations. 
Experimental records of H coverage at known H partial pressure and temperature \cite{EngKui79, WilMatFuk01, MitRosFom03} (\autoref{fig:pressure-conversion}b) are then matched to the \gls{mc} results and used to calculate $\mu^0$, which can fitted by a simple temperature-dependent function.
In \autoref{fig:pressure-conversion}c we show this (linear) fit as well as $\mu^0(T)$ calculated from the NIST-JANAF thermodynamic tables \cite{nist-janaf} and a fit based on the bulk hydride formation from Ref.~\citenum{RahLofFra21}.
The shaded regions correspond to an offset equal to the \gls{rmse} of the linear fit in the \qtyrange{200}{500}{\kelvin} interval.
The NIST-JANAF data is corrected by a constant shift of \qty{-0.2}{\electronvolt} based on the difference between the \gls{ce} H adsorption energy (\qty{-0.63}{\electronvolt}) and experimental reports \cite{LovOls98, WelSilGar15}, ($\sim\qty{-0.43}{\electronvolt}$).
Except for the low-temperature limit, the experimental data is well represented.
It should, however, be noted that most of the experimental data temperatures indicate very low (sub-percent) coverage with H diffusion to the bulk \cite{EngKui79}. 
For this reason, the only data point with substantial coverage at a relevant temperature (highlighted in \autoref{fig:pressure-conversion}a--c) was given a larger weight in the fit.
(The fitting procedure included additional sources of experimental data for other surface orientations, see \autoref{sfig:pressure-conv}.)

\begin{figure*}[t]
\centering
\includegraphics[]{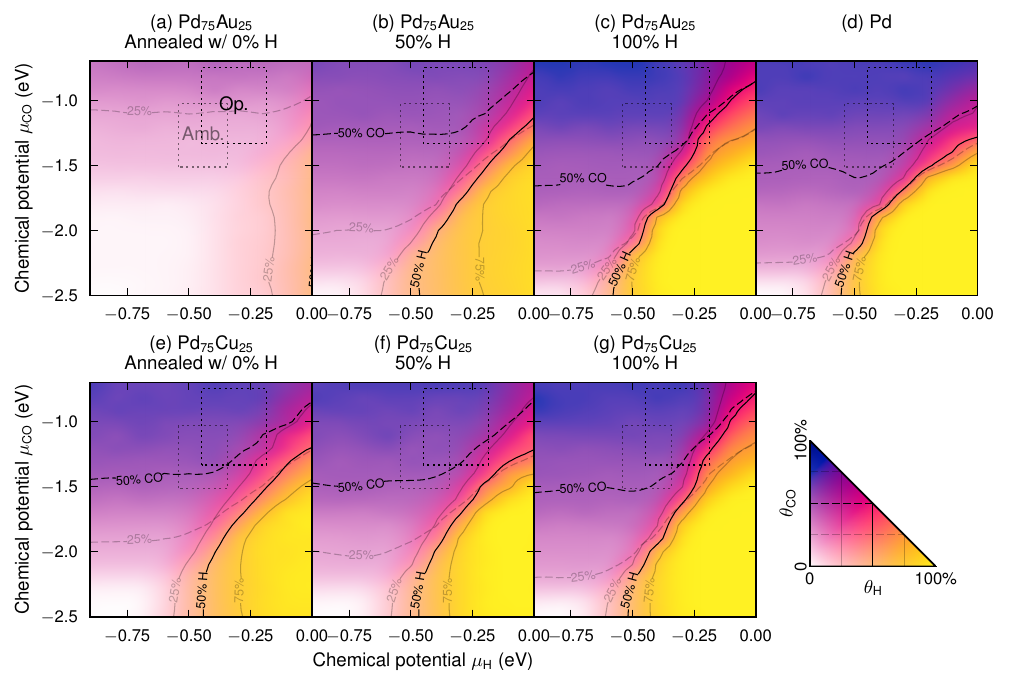}
\caption{
    \textbf{\gls{spd} for Pd, \ce{Pd75Au25} and \ce{Pd75Cu25} prepared in different conditions.}
    The slab configurations were obtained from \gls{mc} simulations at \qty{773}{\kelvin} with (a) 0, (b) 50 or (c) \qty{100}{\percent} H coverage and then covered with CO and H via \gls{mc} simulations at \qty{300}{\kelvin} while keeping the chemical configuration of the slab fixed. 
    The surface coverage is represented by the color according to the colormap in the lower right corner and the contour lines, where the solid lines correspond to H and the dashed to CO. 
    The dotted boxes indicate regions approximately corresponding to operating vs. ambient conditions. 
    }
\label{fig:PdAu-PdCu-SPD}
\end{figure*}

In \autoref{fig:pressure-conversion}d--f the analysis of the pressure conversion procedure is shown for CO. 
The \gls{mc} simulations display a stair-like behavior for the CO coverage, indicating a preference for ordered phases at 33, 50 and \qty{67}{\percent} CO coverage (see \autoref{sfig:Pd-cov} for site-specific coverages) and saturation at \qty{75}{\percent} in line with experimental observations \cite{KaiProBuk03}. 
CO adsorption on Pd has been well-studied for the past decades and experimental reports of the CO coverage over a wide range of pressures and temperatures are available \cite{UntRupFre02, KuhSzaGoo92, KaiProBuk03, HeMemGri88, HeNor88, BehChrErt80}. 
The calculated values for $\mu^0(T)$ have some spread, but overall seem to follow a linear trend that is captured by the fit. 
The calculation based on the NIST-JANAF thermodynamic tables \cite{nist-janaf} include a correction shift of \qty{-0.8}{\electronvolt} based on the approximated difference between the \gls{ce} H adsorption energy ($\sim\qty{-2.3}{\electronvolt}$) and experimental reports \cite{BehChrErt80, SzaKuhGoo93}, ($\sim\qty{-1.5}{\electronvolt}$).
The NIST-JANAF reference differs from the fitted function in slope and offset. 

In \autoref{fig:pressure-conversion}b, e we show how the pressure conversion changes with the different approaches to calculating $\mu^0$. 
Clearly, relatively small differences in $\mu^0$ can lead to shifts of several orders of magnitude in the pressure. 
This finding highlights the inherent difficulty in accurate conversion between chemical potential and partial pressure, rather than a flaw in our methodology.
To account for the uncertainty in partial pressure, we indicate relevant pressure regions in \glspl{spd} below instead of exact partial pressures. 
Due to the lack of experimental data for H, we use all three approaches to determining $\mu^0$ to span the pressure region.
For CO, we rely on the fit with error bars due to the well-corroborated experimental data and lacking treatment of vibrations in the NIST-JANAF approach. 

\subsection{Influence of preparation conditions}

\begin{figure}[t]
\centering
\includegraphics[]{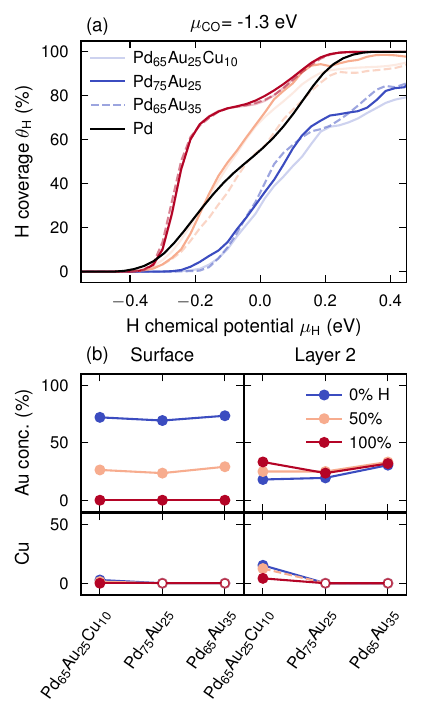}
\caption{
\textbf{H coverage and composition on Pd and PdAuCu-alloys.}
    (a) H coverage as a function of the H chemical potential $\mu_\text{H}$ at fixed CO chemical potential $\mu_\text{CO}=\qty{-1.3}{\electronvolt}$ for Pd, \ce{Pd65Au25Cu10}, \ce{Pd75Au25} and \ce{Pd65Au35} obtained from \gls{mc} simulations at \qty{300}{\kelvin} and with the alloy configuration kept frozen.
    (b) Concentration of the surface layer and subsequent layer for the alloys, obtained from \gls{mc} simulations at \qty{773}{\kelvin} with varying H coverage as indicated by the legend (used also in a). 
}
\label{fig:cov-comp}
\end{figure}

In the following we present \glspl{spd} for Pd-alloy surfaces and study how the preparation conditions affect the H and CO coadsorption.
By preparation conditions we refer to the environment in which the slab configuration equilibrates, which in the present work is set up to mimic the experimental procedure of annealing the samples at \qty{773}{\kelvin} in a \ce{H2} environment \cite{DarNugKad19, DarKhaTom21}. 
Due to the large uncertainty associated with the pressure conversion, we consider a range of fixed H coverages during annealing rather than specifying the \ce{H2} partial pressure via the H chemical potential. 
The resulting compositions of the surface region are presented in \autoref{sfig:surf-comp-111} to \ref{sfig:surf-comp-100}.

We then analyze the coadsorption under operation, i.e. in conditions similar to the experimental measurements \cite{DarNugKad19, DarKhaTom21}, by constructing \glspl{spd} for the H and CO coverage as a function of their respective chemical potentials while keeping the slab configuration fixed. 
The regions corresponding to ambient (\qty{0.6}{\micro\bar} \ce{H2}, \qty{0.1}{\micro\bar} \ce{CO} \cite{NOAJet24}) and operating conditions (\qtyrange{1}{100}{\milli\bar} \ce{H2} \cite{DoE15, DarNugLan21}, \qtyrange{0.1}{5}{\milli\bar} \ce{CO} \cite{DarNugKad19, DarKhaTom21}) are indicated, where the width is associated with the uncertainty in partial pressure conversion rather than the considered pressure intervals. 

\Autoref{fig:PdAu-PdCu-SPD} shows \glspl{spd} for Pd, \ce{Pd75Au25}, and \ce{Pd75Cu25} \{111\}-surfaces prepared by annealing at \qty{773}{\kelvin} with \qty{0}{\percent}, \qty{50}{\percent} or \qty{100}{\percent} H coverage. 
For \ce{Pd75Au25}, the surface coverage clearly depends on the preparation conditions.
Without adsorbed H during annealing, Au segregates to the surface (\autoref{sfig:surf-comp-111}).
This increases the adsorption energy of both CO and H, leading to low coverages (\autoref{fig:PdAu-PdCu-SPD}a) and, in particular, almost no H at the surface under operating conditions which would hinder the \ce{H2} sensing ability. 
With \qty{50}{\percent} H during annealing, the surface composition is similar to the bulk which leads to increased adsorption (\autoref{fig:PdAu-PdCu-SPD}b). 
Lastly, with \qty{100}{\percent} H, the outermost surface layer is made up almost entirely of Pd which leads to a further increase of the overall coverages (\autoref{fig:PdAu-PdCu-SPD}c) resulting in a similar \gls{spd} as to the case of pure Pd (\autoref{fig:PdAu-PdCu-SPD}d). 
There is, however, a notable increase in H coverage in the lower right corner of the operating region, suggesting that \ce{Pd75Au25} with mostly Pd at the surface is superior to Pd in terms of CO adsorption blocking the surface.

For \ce{Pd75Cu25}, the change in H and CO coverages with preparation conditions is less significant (\autoref{fig:PdAu-PdCu-SPD}e--g) and the \gls{spd} falls in between the \qty{50}{\percent} and \qty{100}{\percent} H (coverage during annealing) \glspl{spd} for \ce{Pd75Au25}.
This is because the surface segregation tendency of Cu in PdCu is much weaker than for Au in the case of PdAu, leading to a majority of Pd at the surface under all conditions (\autoref{sfig:surf-comp-111}).
Based on the H coverage contour lines in the vicinity of the operating region, \ce{Pd75Cu25} allows for slightly more H at the surface compared to Pd. 

In summary, alloys containing Au are greatly influenced by the preparation conditions in terms of their ability to adsorb H. 
\Autoref{fig:cov-comp} shows that the H adsorption as a function of the H chemical potential shifts by at least \qty{0.2}{\electronvolt} with the annealing H coverage, corresponding to several orders of magnitude for the pressure (\autoref{fig:pressure-conversion}a--b). 
\Autoref{fig:cov-comp} also indicates that the alloy composition plays a secondary role, as will be further discussed in the next section.

\subsection{Influence of alloy composition}

\begin{figure}
\centering
\includegraphics{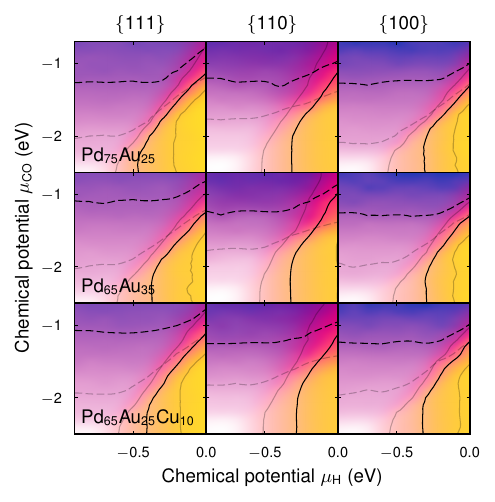}
\caption{
    \textbf{Composition and orientation-dependent \glspl{spd}.}
    \Gls{spd} for \ce{Pd75Au25}, \ce{Pd65Au35} and \ce{Pd65Au35Cu10} at varying H and CO chemical potentials obtained from \gls{mc} simulations at \qty{300}{\kelvin} with alloy configuration kept frozen.
    The slab configurations are annealed at \qty{773}{\kelvin} with \qty{50}{\percent} H coverage.
    The coverages follow the same colormap as \autoref{fig:PdAu-PdCu-SPD}, with solid and dashed contour lines for H and CO, respectively.
    Note that for \{110\}, \qty{100}{\percent} coverage corresponds to 2 ML of adsorbates due to the large surface area per metal atom. 
}
\label{fig:PdAuCu-spd}
\end{figure}

\begin{figure*}
\centering
\includegraphics{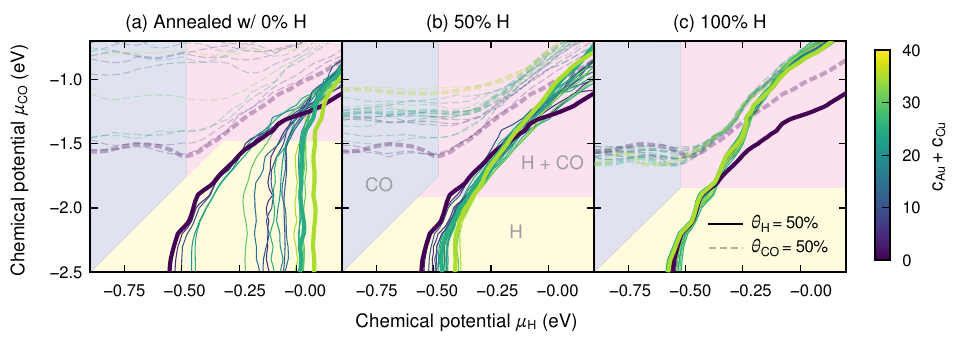}
\caption{
    \textbf{50\% coverage contour lines for Pd and various PdAuCu alloy \{111\} surfaces.}
    The slab configurations are obtained from \gls{mc} simulations at \qty{773}{\kelvin} with (a) \qty{0}{\percent}, (b) \qty{50}{\percent}, and (c) \qty{100}{\percent} H coverage.
    The H and CO coverages are obtained from subsequent \gls{mc} simulations at \qty{300}{\kelvin} with fixed slab configuration.
    All alloy compositions considered are represented, with Pd, \ce{Pd75Au25}, and \ce{Pd65Au25Cu10} highlighted by thicker lines (see \autoref{sfig:surf-comp-111} for details concerning the compositions).
    The colored region indicate the limits of H, CO and H--CO adsorption.
}
\label{fig:PdAuCu-contour-lines}
\end{figure*}

\Autoref{fig:PdAuCu-spd}a shows \glspl{spd} for \ce{Pd75Au25}, \ce{Pd65Au35}, and \ce{Pd65Au35Cu10} prepared with \qty{50}{\percent} H coverage (the upper right corner corresponds to \autoref{fig:PdAu-PdCu-SPD}b). 
As noted also in \autoref{fig:cov-comp}a, the CO and H coadsorption behavior is seemingly independent of the alloy composition. 
In particular, there is no distinct effect of introducing Cu compared to Au, which is unexpected given the strong experimental evidence for Cu, specifically, reducing CO poisoning \cite{DarNugKad19, DarKhaTom21}.
This (lack of a) trend remains across alloy compositions and preparation conditions (\autoref{sfig:SPD-all-111}--\ref{sfig:SPD-all-100}). 

\Autoref{fig:PdAuCu-contour-lines} shows how the \qty{50}{\percent} H adsorption contour line (from the \glspl{spd}) shifts with alloy composition, which can be interpreted as an indicator for the resistance to CO poisoning (with respect to adsorption thermodynamics).
Three regions can be identified: the H adsorption limit at the bottom, the CO adsorption limit to the left, and the H--CO coadsorption limit in the upper right corner. 
In the H adsorption limit (lower right quadrant)., Pd generally adsorbs more H than the alloys and the contour lines move to higher H chemical potentials with decreasing Pd content. 
A similar trend can be identified for CO, in the CO adsorption limit (upper left quadrant).

In the coadsorption limit (upper right quadrant), however, we see a shift in the trend where the \qty{50}{\percent} H contour line for Pd crosses the corresponding contour lines for the alloys, meaning that the alloys adsorb more H at a given chemical potential. 
This suggests that in the coadsorption limit, alloying, in general, is beneficial for CO poisoning reduction.
Depending on the preparation condition, this crossing of the contour lines happens inside (\qty{100}{\percent} H) or outside (\qty{0}{\percent} H) of the operating region ($\mu_\text{H} < \qty{-0.2}{\electronvolt}$).
This suggests that a high H coverage during preparation is necessary to achieve the CO poisoning mitigating effects.
We emphasize, however, again that there is a large uncertainty associated with the pressure conversion, and these predictions should be viewed as qualitative rather than quantitative.   

For the systems prepared in \qty{100}{\percent} H, the spread in the contour lines is generally narrow, which is reasonable since all systems have close to \qty{100}{\percent} Pd at the surface (\autoref{sfig:surf-comp-111}).
In the coadsorption limit, however, all contour lines for the alloys distinctly deviate from the contour lines for Pd. 
This is surprising since their concentrations vary over a wide range \emph{except} for the surface layer, which is nearly identical to the pure Pd case. 
It thus appears that any amount of alloyant in the bulk and subsurface region can distinctly change the adsorption behavior without entering the surface layer and without any dependence on the concentration. 

\begin{figure}
\centering
\includegraphics{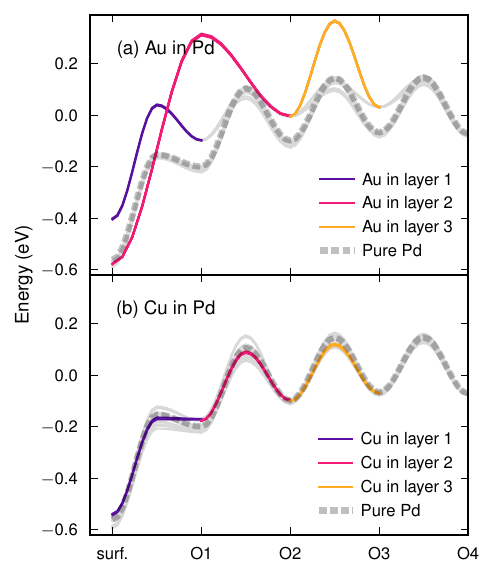}
\caption{
    \textbf{Kinetic barriers for dilute Pd-alloys.}
    Migration barriers for a H atom moving from the surface into the material via octahedral absorption sites, for pure Pd (dashed) and structures with a single (a) Au or (b) Cu atom placed in one of the three outermost atomic layers i.e., the surface region).
    The paths closest to the Au/Cu atom are highlighted by the colors indicated by the legend. 
    Note that for the case of Au in layer 2, the O1 site is not stable and the H atom relaxes to the surface. 
}
\label{fig:PdAu-PdCu-barriers}
\end{figure}

While these findings explain why alloying might mitigate CO poisoning, they do not provide a rationale for the benefit of adding Cu, for which there is strong experimental evidence \cite{DarNugKad19, DarKhaTom21}.
This suggests that \emph{adsorption thermodynamics} are not enough to fully understand CO poisoning.
Given the results above, we can hypothesize that under operating conditions, CO blocks the surface entirely for Pd and only partially for the alloys. 
The unoccupied areas will correspond to less favorable adsorption sites, i.e., less Pd than in the covered regions. 
It is then possible that the presence of Cu, especially in the second atomic layer, might facilitate the sorption of H into the bulk. 
To investigate this, we look at migration paths and associated kinetic barriers via \gls{neb} calculations using the MACE model.

We consider \{111\} surfaces consisting of Pd with a single Au or Cu atom placed in atomic layer 1, 2, or 3, with a single H in the \gls{fcc} adsorption site or one of the octahedral absorption sites, and mapped out all paths through the slab along with their associated kinetic barriers (\autoref{fig:PdAu-PdCu-barriers}). 
For PdAu, paths close to the Au atom are associated with a large increase in the energy of the initial and final states as well as the barrier. 
For PdCu, on the other hand, the migration energetics remain mostly unchanged or slightly improved compared to Pd. 
This suggests that the presence of Cu in the surface region could improve the \emph{sorption kinetics}, especially in comparison to Au, and thereby enable H sorption in situations where the most favorable paths through the surface are blocked by CO.

At first glance, this negative impact of Au on H sorption appears to contradict prior studies, which demonstrate that Au enhances sorption kinetics in Pd-based alloys \cite{WadNugLid15, NamOguOhn18}.
We argue, however, that while Au \emph{globally} improves sorption kinetics, through lattice expansion and a reduction in the total energy barrier for H sorption (mainly by increasing the adsorption energy) \cite{NamOguOhn18}, it \emph{locally} inhibits the kinetics for paths that passes an Au atom. 
This local effect becomes critical when the most energetically favorable H sorption paths are blocked by CO.

\subsection{Influence of surface orientation}

In the discussion above, we have mainly focused on the \{111\} surface, since it is the minimum energy surface for all three alloyants \cite{TraXuRad16}.
\Autoref{fig:PdAuCu-spd} shows \glspl{spd} for the \{110\} and \{100\}, and additional results for the other surface orientation are available in the Supporting Information (\autoref{sfig:surf-comp-110} -- \ref{sfig:contour-lines-all}).
There are some minor changes of the \glspl{spd} between the orientations, most notable for \{110\} compared to the other surfaces, but overall the conclusions from the previous sections hold, namely that the coadsorption behavior depends primarily on the preparation conditions.

\sloppy Some orientation-dependent effects can be pointed out.
\{110\} alloy surfaces show less benefit compared to Pd in the coadsorption limit, in terms of the ability to adsorb H (\autoref{sfig:contour-lines-all}).
In addition, the surface segregation behavior of \{110\} PdCu, specifically, differs from the other orientations in so far that Cu segregates to the surface in vacuum (\autoref{sfig:surf-comp-110}) to a much larger degree compared to the other orientations. 
\{100\} alloy surfaces display very similar trends to \{111\} with a slightly larger benefit compared to pure Pd in the coadsorption limit  (\autoref{sfig:contour-lines-all}).
In the lower right corner of what we consider operation conditions ($\mu_\text{H}=\qty{-0.2}{\electronvolt}$, $\mu_\text{CO}=\qty{-1.3}{\electronvolt}$), we find an increase in H adsorption with Cu concentration for the alloy prepared in \qty{100}{\percent} H which is the only clear example of Cu positively impacting CO poisoning characteristics (\autoref{sfig:cov-0.2-1.3}).

\section{Conclusions}

In the present work, we have studied the coadsorption of H and CO on PdAuCu alloy surfaces with \qtyrange{0}{35}{\percent} Au and \qtyrange{0}{25}{\percent} Cu to understand the mitigating effect on CO poisoning of such alloys, compared to Pd, in the context of \ce{H2} sensing.
We have found that by tuning the surface composition via the preparation conditions, specifically the H coverage, the adsorption of H and CO during operation can be tuned.
Increasing the amount of Au at the surface, which can be achieved by preparing the surfaces at low H coverages, significantly reduces the CO adsorption tendency. 
Unfortunately, this also reduces the H adsorption such that higher H partial pressures are necessary to adsorb a comparable amount of H. 
On the other hand, when H coverage is high during preparation, one obtains Pd-rich surfaces that are prone to adsorb both H and CO, but with a higher H to CO ratio than pure Pd. 
This suggests that PdAuCu alloys prepared in a H-rich environment lead to improved CO poisoning resistance based on \emph{adsorption thermodynamics}.

\begin{figure}
\centering
\includegraphics[width=0.5\textwidth]{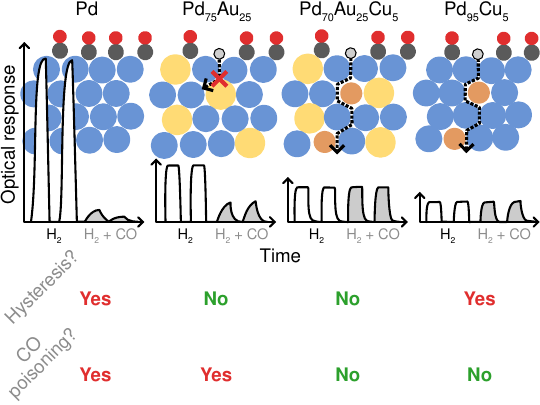}
\caption{
    \textbf{Proposed mechanism for Cu-induced CO poisoning resistance.}
    The upper panel shows the (absolute) shift in optical response of Pd, \ce{Pd75Au25}, \ce{Pd70Au25Cu5}, and \ce{Pd95Cu5} when subjected to pulses of \ce{H2} (\qty{40}{\milli\bar}) vs. \ce{H2} (\qty{40}{\milli\bar}) + \ce{CO} (\qty{5}{\milli\bar}), adapted from \cite{DarNugKad19}.
    The slab representations show the corresponding proposed slab--CO--H interplay, where \ce{Pd} is completely blocked by CO, \ce{Pd75Au25} suffers from poor kinetics due to Au in the surface region, while alloys with Cu are able to shuttle H into the material via Cu in the surface region. 
    The lower panel summarizes the corresponding hysteresis and CO poisoning characteristics. 
}
\label{fig:CO-poisoning}
\end{figure}

Adsorption thermodynamics alone are, however, not sufficient to explain the experimentally observed mitigating effect of Cu on CO poisoning \cite{DarNugKad19, DarKhaTom21}. 
The experimentally observed optical shifts in response to \ce{H2} vs. a mixture of \ce{H2} and CO (\autoref{fig:CO-poisoning}), reveal that the primary difference between the samples during CO exposure is not the absolute H sorption, but rather the relative H sorption compared to the CO-free case and, importantly, the \emph{sorption kinetics}. 

With regard to the kinetic aspect, we have shown that in the dilute limit, introducing Au in the surface region of a Pd slab results in a significant increase in the kinetic barriers associated with H absorption, while Cu does not affect the H absorption energetics compared to pure Pd. 
Based on this finding, we suggest the following rationale for the observed CO poisoning behavior of the different systems (\autoref{fig:CO-poisoning}). 
For pure Pd, the surface is almost completely blocked by CO, resulting in slow sorption of small amounts of H. 
For the alloys, the surface is only partially covered by CO, leading to some H adsorption. 
Crucially, while Au in the near-surface hinders H sorption, Cu has no pronounced effect on the sorption kinetics compared to Pd. 
We thus suggest that the introduction of Cu creates viable pathways for H sorption in cases where the most favorable pathways are blocked by CO at the surface.

Further studies of the bulk absorption process in realistic systems are necessary to further support the suggested mechanism regarding the Cu-induced improvement in H sorption kinetics. 
The \glspl{mlip} developed in the present work for the PdAuCu:(H,CO) system provide an ideal starting point for this endeavor.

\section*{Acknowledgments}

We gratefully acknowledge funding from the Swedish Research Council (Nos.~2020-04935 and 2021-05072) and the Area of Advance at Nano at Chalmers as well as computational resources provided by the National Academic Infrastructure for Supercomputing in Sweden at NSC, PDC, and C3SE partially funded by the Swedish Research Council through grant agreement No.~2022-06725, as well as the Berzelius resource provided by the Knut and Alice Wallenberg Foundation at NSC, as well as the Berzelius resource provided by the Knut and Alice Wallenberg Foundation at NSC.

\section*{Data Availability}

The \gls{nep} and MACE models as well as the database of \gls{dft} calculations used to train these models have been deposited in the Zenodo database under accession code \href{https://doi.org/10.5281/zenodo.17670909}{10.5281/zenodo.17670908}.

\end{document}